\documentclass[aps,twocolumn,
showkeys,showpacs,nofootinbib,byrevtex]{revtex4-1}
\usepackage{amsmath,amsfonts,amssymb,amscd,amsxtra,amsthm}
\usepackage{graphicx}
\usepackage{bm}
\usepackage{epstopdf}
\begin{document}

\title{Reggeized model for $\gamma p \to \rho^- \Delta^{++}(1232)$
photoproduction}
\author{Byung-Geel Yu}
\email[E-mail: ]{bgyu@kau.ac.kr}
\affiliation{Research Institute of Basic Sciences,
Korea Aerospace University, Goyang, 10540, Korea}
\author{Kook-Jin Kong}
\email[E-mail: ]{kong@kau.ac.kr}
\affiliation{Research Institute of Basic Sciences,
Korea Aerospace University, Goyang, 10540, Korea}
\date{\today}

\begin{abstract}

A model for the  reaction $\gamma p\to\rho^-\Delta^{++}$ is
presented with the $t$-channel $\pi+\rho$ exchanges reggeized to
describe the reaction up to high energies. Gauge invariance of
$\rho$ exchange is discussed in connection with the convergence of
the reaction cross section at high energy. The roles of
electromagnetic (EM) multipole moments of $\Delta^{++}$ baryon and of
$\rho^-$ meson are analyzed in total and differential cross sections
and spin density matrix elements. Photon polarization asymmetry $\Sigma$ is
predicted for a measurement of electromagnetic moments of $\rho^-$
and of $\Delta^{++}$.
\end{abstract}

\pacs{11.55.Jy, 13.60.Rj, 13.60.Le, 13.85.Fb, 14.20.Gk, 14.40.Be}
\keywords{$\Delta(1232)$ and $\rho$ photoproduction, Gauge
invariance Electromagnetic multipole moments, Regge trajectory}

\maketitle

\section{introduction}

Photoproduction of charged $\rho$ with $\Delta$ baryon in the
final state is one of the issues which could offer an opportunity
to study the interaction between hadrons of higher spins.
Moreover, since these particles are fundamental entities to
constitute the $\pi N$ system, to investigate electromagnetic
properties of $\rho$  and of $\Delta$ as well in the reaction
provides information on the internal excitation of $\pi\pi$ and
$\pi N$ systems through their EM multipole moments \cite{bgyu-rho}
\cite{bgyu-pi-delta,bgyu-pi-delta2}.

On the other hand, in relation with recent topics on the GlueX
experiments \cite{ghoul}, a theoretical framework for the reaction
$\gamma p\to\rho^-\Delta^{++}$ will be a tool useful for the
analysis of multipion photoproductions focusing on the search of
exotic mesons such as $\pi_1(1400)$ and $\pi_1(1600)$ of spin-1
particles \cite{szczepaniak,schott,nys}. As the gluon excitation
in the $q\bar{q}$ structure of mesons may hint at our
understanding of quark confinement, study of the reaction $\gamma
p\to\pi_1\Delta$ is of significance, and, hence, modelling the
$\gamma p\to\rho\Delta$ process and verifying its reliability by
existing data are meaningful, prior to application. To date,
however, few theoretical studies on the reaction are found in
literature, though energy and angle dependence of reaction cross
section had been measued in various photon energies
\cite{barber}\cite{abramson}\cite{struczinski,nelson,eisenberg,ballam}.
This is mainly because of the difficulties in handling divergences
of $\rho$ and $\Delta$ particles at high energies.

With the empirical evidence for $\rho\Delta$ formation in the
final state from the analysis of the reaction process $\gamma p\to
p\pi^+\pi^0\pi^-$ \cite{erbe},
a theoretical study of the $\gamma p\to \rho^-\Delta^{++}$ process
was attempted in Ref. \cite{clark-rho-delta}, but the results were
limited only to a discussion of an ad hoc prescription for
gauge invariance of the $\rho$ exchange with EM multipoles.
The analysis of the $\gamma p\to\rho^-\Delta^{++}$
process by using the $\pi+b_1+\rho+a_2$ Regge-poles to fit to data
in the $s$-channel helicity amplitude was presented in Ref.
\cite{mclark}. A qualitative calculation of the reaction  $\gamma
p\to\rho^\pm\Delta$ was performed for a subprocess in the $\pi N$
electroproduction \cite{laget}. However, all these were not
complete to fully illustrate the production mechanism of
$\rho$ meson and $\Delta$ baryon  with their EM properties from
the standpoint of the effective Lagrangian formulation.

In previous works we investigated  photoproduction of spin-1
vector meson on nucleon, $\gamma N\to\rho^\pm N$ \cite{bgyu-rho},
and of spin-3/2 $\Delta$ baryon, $\gamma p\to \pi^\pm\Delta$
\cite{bgyu-pi-delta,bgyu-pi-delta2} by using the relativistic Born
amplitudes, where the $t$-channel meson exchanges $\pi+\rho$, and
$a_2$ exchange further in the latter reaction, were reggeized
following the procedure of Ref. \cite{bgyu-regge}. In both
reactions the energy-dependence of cross sections exhibited a
feature of the nondiffractive two-body scattering with a steep
decrease over the resonance region as photon energy increases.
From the viewpoint of the Regge formalism, this can be understood
by the relation, $\sigma\sim s^{\alpha_J(0)-1}$, which predicts
the dominance of the $\pi$ exchange over the $\rho$ in the total
cross section.
Indeed, our previous analyses on these photoproduction processes
showed an agreement with  such a production mechanism without
either fit-parameters \cite{mclark}, or ad hoc counter terms
\cite{clark-rho-delta}. On the other hand, both the reactions
$\gamma p\to\pi^-\Delta^{++}$ and $\gamma N\to\rho^\pm N$ have the
terms proportional to $p^2/m^2$ in their respective propagators
which are known to diverge at high energy. For the convergence of
reaction cross sections, the former reaction needed a special
gauge prescription, i.e., the minimal gauge prescription which renders
only the charge coupling $\gamma\Delta\Delta$ interaction to
contribute at high energies \cite{stichel,clark-pi-delta}. In
the latter case, however, the cross section showed a smooth convergence
without any constraints on the $\rho$ meson EM multipoles further,
for the reason as discussed briefly in the appendix.
Recalling that the contribution of the $u$-channel baryon pole is,
in general, limited to a lower energy region, whereas that of
$\rho$ exchange proceeds up to high energies via the peripheral
scattering in the $t$-channel, it seems natural to remove the
transverse component of $\gamma\Delta\Delta$ including EM dipole
and quadrupole moments of $\Delta^{++}$ baryon so that they could
give no more contributions at high energies. Instead, the role of
$\rho^\pm$ meson magnetic dipole and electric quadrupole moments
was found to be less significant in comparison to those of
$\Delta^{++}$ baryon in each reaction process \cite{bgyu-pi-delta2}.
In establishing such convergent theories, the Ward
identity at the $\gamma\Delta\Delta$, and at the $\gamma\rho\rho$
vertex, respectively, plays a crucial role to dictate a correct form of
charge coupling, as pointed out in
Refs. \cite{bgyu-pi-delta,bgyu-rho}.

Hinted by these findings, we here investigate the $\gamma
p\to\rho^-\Delta^{++}$ process as an extension of our
previous works, and our theoretical interest in the present issue
is two-folded; the respective roles of the  EM multipole moments of
$\rho^-$ meson and $\Delta^{++}$ baryon in the reaction, and convergence of the
reaction cross section at high energy, because both the propagators
of the spin-1 and spin-3/2 particles have the divergent term
which needs a special care.

\section{The Reggeized amplitude}

Since a particle of spin-$J$ has $2J+1$ EM multipole moments,
there are four EM multipole moments at the $\gamma\Delta\Delta$
vertex in addition to the three multipole moments at the
$\gamma\rho\rho$ vertex in the charged process $\gamma p\to
\rho^-\Delta^{++}$.  As
discussed in Ref. \cite{bgyu-rho}, the validity of the Ward
identity for the longitudinal component of the $\gamma\rho\rho$ coupling
is crucial to render gauge invariance the more simple form.
This is true for the
$\gamma\Delta\Delta$ vertex as well \cite{bgyu-pi-delta}. In this
work, therefore, based on the Ward identities at these charge
couplings vertices we first formulate a gauge invariant
model for the $\gamma p\to \rho^-\Delta^{++}$ process with the
full propagators taken into account for the $\Delta$ baryon
and $\rho$ meson. This will make the model description complete
for the interaction between higher-spin particles with the EM
multipole moments fully considered. We, then, proceed to
higher energy region to investigate  convergence of the
reaction cross section.

\begin{figure}[]
\includegraphics[width=5cm]{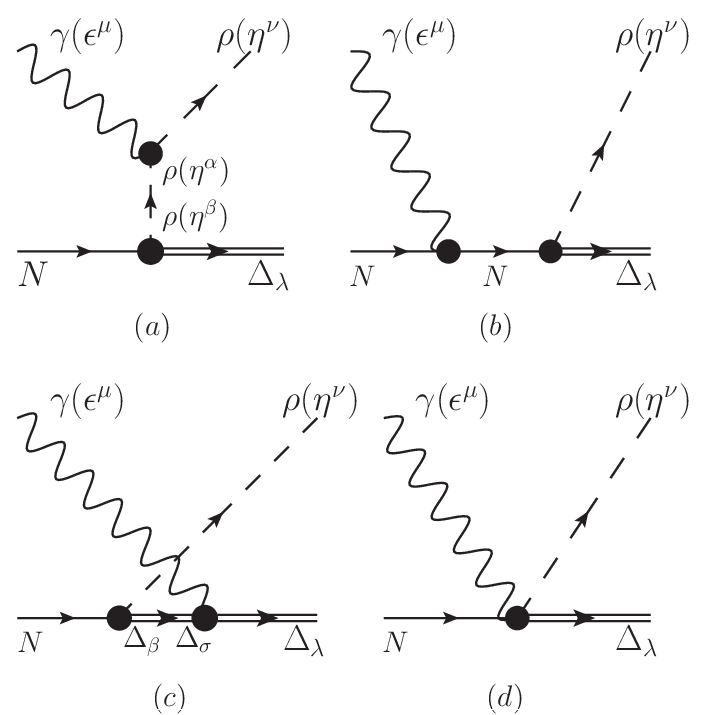}
\caption{Feynman diagrams for the gauge-invariant $\rho$ exchange.
(a), (b), (c), and (d) depict each term in Eq. (\ref{born})
in order.
$\pi$ exchange proceed via the $t$-channel exchange in (a).}
\label{fig1}
\end{figure}

\subsection{Conventional approach to gauge invariance: Model I}

For the photoproduction process
\begin{eqnarray}
\gamma(k) +p(p)\to \rho^-(q) +\Delta^{++}(p'),
\end{eqnarray}
we denote the momenta of the initial photon, proton and the final
$\rho^-$ and $\Delta^{++}$ by $k$, $p$, $q$, and $p'$,
respectively. The charge conservation, $e_p-e_{\rho^-}-e_{\Delta^{++}}=0$,
requires inclusion of $\Delta^{++}$-pole in the $u$-channel
in addition to the $s$-channel proton-pole for the $t$-channel
$\rho^-$ exchange to be conserved. Thus, we write the Born amplitude
for the gauge-invariant $\rho$ exchange in the $t$-channel as
\begin{eqnarray}\label{born}
&&M_\rho=\overline{u}_{\lambda}(p')\eta^*_\nu(q)\nonumber\\&&
\hspace{1cm}\times
\left[{M}^{\lambda\nu\mu}_{t(\rho)}+M^{\lambda\nu\mu}_{s(N)}+
M^{\lambda\nu\mu}_{u(\Delta)}+ {M}^{\lambda\nu\mu}_c\right] \epsilon_\mu(k)u(p).
\ \ \
\end{eqnarray}
in terms of nucleon and $\Delta$-poles for the sake of generality.
Here $M_c$ is the contact interaction term which is usually called for
when the meson-baryon coupling is of the derivative type.
$u_\lambda(p')$, $u(p)$, $\eta_\nu(q)$, and $\epsilon_\mu(k)$
are the spin-3/2 $\Delta^{++}$ spinor of the Rarita-Schwinger field,
Dirac spinor for nucleon, and the spin polarizations of $\rho$ and
photon, respectively.
In Eq. (\ref{born}), nucleon, $\rho$, and $\Delta$ exchanges in the
respective channels are given by
\begin{eqnarray}
&&{M}^{\lambda\nu\mu}_{s(N)}= \Gamma_{\rho
N\Delta}^{\lambda\nu}(q,p',p+k)
\frac{\rlap{\,/}p+\rlap{\,/}k+M_N}{s-M^{2}_{N}}\Gamma^\mu_{\gamma NN}(k),
\label{nucleon}\\
&& {M}^{\lambda\nu\mu}_{t(\rho)}
=\Gamma^{\nu\mu\alpha}_{\gamma\rho\rho}(q,Q)
\frac{D^\rho_{\alpha\beta}(Q)}{t-m^{2}_{\rho}} \Gamma_{\rho
N\Delta}^{\beta\lambda}(Q,p',p),
\label{rho}\\
&&{M}^{\lambda\nu\mu}_{u(\Delta)}=
\Gamma_{\gamma\Delta\Delta}^{\lambda\mu\sigma}(k)\frac{\rlap{/}p'-\rlap{\,/}k+M_\Delta}
{u-M^{2}_{\Delta}}\Pi^\Delta_{\sigma\beta}(p'-k)\nonumber\\&&\hspace{1.5cm}\times
\Gamma_{\rho N\Delta}^{\beta\nu}(q,p'-k,p)\ \ \label{delta}
\end{eqnarray}
with the $t$-channel momentum transfer $Q^\mu=(q-k)^\mu$, and
the spin projection operators
\begin{eqnarray}
&&D^\rho_{\alpha\beta}(p)=-g_{\alpha\beta} +{p_\alpha p_\beta\over
m^2_\rho}\,,\\ \label{rho-propagator}
&&\Pi^\Delta_{\mu\nu}(p)=-g_{\mu\nu}+\frac{\gamma_{\mu}\gamma_{\nu}}{3}
+\frac{\gamma_{\mu}p_{\nu}-\gamma_{\nu}p_{\mu}}{3M_{\Delta}}
+\frac{2p_{\mu}p_{\nu}}{3M^{2}_{\Delta}}\,,\ \ \label{delta-propagator}
\end{eqnarray}
for spin-1 and spin-3/2, respectively.
The EM coupling vertices $\gamma NN$, $\gamma\rho\rho$
\cite{bgyu-rho}, and
$\gamma\Delta\Delta$ \cite{bgyu-pi-delta} which fully account for
their EM multipole moments are defined as follows,
\begin{eqnarray}
&&\epsilon_\mu\Gamma^\mu_{\gamma
NN}(k)=e_N\rlap{/}\epsilon-\frac{e\kappa_N}{4M_N}\left[\rlap{/}\epsilon,\,
\rlap{\,/}k\right],\label{gammann}\\
&&\epsilon_\mu\Gamma_{\gamma\Delta\Delta}^{\lambda\mu\sigma}(p',k,p)=
    -\biggl\{e_\Delta(g^{\lambda\sigma}\rlap{/}\epsilon-\epsilon^\lambda\gamma^\sigma
    -\gamma^\lambda \epsilon^\sigma +\gamma^\lambda\rlap{/}\epsilon
    \gamma^\sigma)\nonumber\\
    &&\hspace{2cm}-{e\over4M_\Delta}\left( \kappa_\Delta\,g^{\lambda\sigma}
    +\chi_\Delta{k^\lambda
        k^\sigma\over4M^2_\Delta}\right)\left[\rlap{/}{\epsilon},\,\rlap{\,/}{k}\right]
    \nonumber\\&&\hspace{2cm}
    +{e\lambda_\Delta\over
        4M^2_\Delta}\left[k^\lambda k^\sigma
    \rlap{/}\epsilon-{1\over2}\rlap{\,/}k(\epsilon^\lambda
    k^\sigma+\epsilon^\sigma k^\lambda) \right]\biggr\},\label{gamma4}\\
&&\epsilon_\mu\Gamma^{\nu\mu\alpha}_{\gamma\rho\rho}(q,Q)=-e_\rho\biggl\{
\biggl[(q+Q)^\mu g^{\nu\alpha}-Q^\nu g^{\mu\alpha}-q^\alpha
g^{\mu\nu}\biggr]\nonumber\\&&\hspace{2.5cm}
+\kappa_\rho(g^{\mu\alpha}k^\nu- g^{\mu\nu}k^\alpha)\nonumber\\
&&\hspace{2.5cm} -{(\lambda_\rho+\kappa_\rho)\over
2m_\rho^2}\biggl[(q+Q)^\mu k^\nu k^\alpha
\nonumber\\&&\hspace{2.5cm} -{1\over2}(q+Q)\cdot k(k^\nu
g^{\mu\alpha}+k^\alpha
g^{\mu\nu})\biggr]\biggr\}\epsilon_\mu.\label{gvv}
\end{eqnarray}
Here $e_\Delta$, $\kappa_\Delta$, $\chi_\Delta$, and
$\lambda_\Delta$  are charge, anomalous magnetic moment,
magnetic octupole, and electric quadrupole moments of the
$\Delta$, respectively.  $e_\rho$, $\kappa_\rho$ and
$\lambda_\rho$ are  charge, anomalous magnetic moment and
electric quadrupole moment of $\rho$ meson. In specfic, $e_p=e$,
$e_{\rho^-}=-e$, and $e_{\Delta^{++}}=2e$, respectively.

Note that all the terms excluding the charge coupling terms in Eqs.
(\ref{gammann}), (\ref{gamma4}), and (\ref{gvv}) are gauge
invariant by themselves, while those charge coupling terms
satisfy the Ward identities in their respective vertices
\cite{bgyu-rho,bgyu-pi-delta}.

For the strong coupling vertex $\rho N\Delta$ we utilized the
following form  \cite{clark-rho-delta}
\begin{eqnarray}\label{rhondelta}
&&\Gamma^{\lambda\nu}_{\rho N\Delta}(q,p',p)=\bigg[{f_{\rho
N\Delta}\over m_\rho}\left( q^\lambda\gamma^\nu - \rlap{/}q
g^{\lambda\nu} \right)\nonumber\\&&\hspace{2cm}+{g_{\rho
N\Delta}\over m^2_\rho}\left(q^\lambda p'^\nu -q\cdot p'
g^{\lambda\nu} \right)\nonumber\\&&\hspace{2cm}+{h_{\rho
N\Delta}\over m^2_\rho} \left(q^\lambda p^\nu -q\cdot p\,
g^{\lambda\nu} \right) \bigg]\gamma_5
\end{eqnarray}
with the quark model prediction, $f_{\rho^-
p\Delta^{++}}={6\sqrt{2}\over5}f_{\rho^0 pp}=8.57$, which is from
the relation ${f_{\rho^0 pp}\over m_\rho}={g_{\rho^0
pp}\over2M}(1+K_\rho)$  \cite{gebrown}. From the dispersion
relation for $\pi\pi\to N\overline{N}$, the $\rho NN$ coupling
constants  were determined to be $f_\rho^2/4\pi= 2.01\sim 3.34$
and $\kappa_\rho=3.5\sim 6.6$ \cite{hoehler}. Within the present
framework $f_\rho=2g_{\rho NN}=5.2$ and $\kappa_\rho=3.7$ are best
to describe hadron reactions including $\rho$ exchange
\cite{yu-kong-pin}. On the other hand, the universality of $\rho$
meson coupling to hadrons implicates $f_{\rho^0 p\Delta^+}=
2g_{\rho^0 pp}$ and $f_{\rho^-p\Delta^{++}}=\sqrt{3/2}f_{\rho^0
p\Delta^+}=6.37$ by the isospin relation. In numerical analyses
the $f_{\rho^0 p\Delta^{+}}$ was determined to be in the range
$4.91 - 7.81$ from a mesonic model \cite{liu} and $3.5 - 7.8$ from
other hadronic process \cite{kammano}. Among the values discussed
above we choose $f_{\rho^- p\Delta^{++}}$ which properly
reproduces experimental data.

For consistency with the previous work on $\gamma N\to\rho^\pm
N$ the EM multipole moments of $\rho$ meson, $\kappa_\rho$ and
$\lambda_\rho$, are taken the same as in Ref. \cite{bgyu-rho}. The
values for the EM multipole moments of $\Delta^{++}$ baryon in Eq.
(\ref{gamma4}) are resumed from Ref. \cite{bgyu-pi-delta} for a complete set
of these observables.

With these together in Eq. (\ref{born}), current conservation of
the Born amplitude leads to the contact interaction of the form,
\begin{eqnarray}\label{contact}
&&{M}_c=-\overline{u}^\lambda(p')\biggl\{e_\rho\frac{f_{\rho
N\Delta}}{m_\rho}(\epsilon_\lambda\rlap{/}{\eta^*}
-\eta^*_\lambda\rlap{/}{\epsilon})\nonumber\\
&&+\frac{g_{\rho
N\Delta}}{m_\rho^2}\left[e_\Delta\left(q_\lambda\epsilon\cdot{\eta^*}
-\eta^*_\lambda{\epsilon}\cdot
q\right)+e_\rho\left(\epsilon_\lambda p'\cdot\eta^*
-\eta^*_\lambda\epsilon\cdot p'\right)\right]\nonumber\\
&&+\frac{h_{\rho
N\Delta}}{m_\rho^2}\left[e_N\left(q_\lambda\epsilon\cdot{\eta^*}
-\eta^*_\lambda{\epsilon}\cdot
q\right)+e_\rho\left(\epsilon_\lambda
p\cdot\eta^*-\eta^*_\lambda\epsilon\cdot
p\right)\right]\biggr\}\nonumber\\&&\times\gamma_5 u(p),
\end{eqnarray}
which deserves comparison to the well-known case of $\pi^-\Delta^{++}$
photoproduction \cite{bgyu-pi-delta}.

Following the conventional recipe for reggeization, the
Reggeized $\rho$ exchange is now written as,
\begin{eqnarray}\label{rho-regge}
{\cal M}_\rho= M_\rho \times(t-m_\rho^2){\cal R}^\rho(s,t),
\end{eqnarray}
where
\begin{eqnarray}\label{regge1}
{\cal R}^\varphi(s,t) =\frac{\pi\alpha'_\varphi\times {\rm phase}
}{\Gamma(\alpha_\varphi(t)+1-J)\sin\pi\alpha_\varphi(t)}
\left(\frac{s}{s_0}\right)^{\alpha_\varphi(t)-J}
\end{eqnarray}
is the Regge-pole written collectively for a meson $\varphi$ of
spin-$J$  with the  phase, trajectory $\alpha(t)_J$, and $s_0$=1
GeV$^2$.

The gauge-invariant pion exchange in the $t$-channel is given by,
\begin{eqnarray}\label{amp2}
&&{\cal M}_{\pi}=-i{g_{\gamma\pi\rho}\over m_0}{f_{\pi
N\Delta}\over m_\pi}\, \varepsilon_{\mu\nu\alpha\beta}\epsilon^\mu
\eta^{*\nu} k^\alpha Q^\beta
\overline{u}_\lambda(p')Q^\lambda u(p)\nonumber\\
&&\hspace{1.5cm}\times{\cal R}^\pi(s,t),
\end{eqnarray}
where $g_{\gamma\pi\rho}=\pm0.224$ is estimated from the measured
decay width and $f_{\pi^-p\Delta^{++}}$ is chosen in the range
from $1.7$ to $2.16$ \cite{bgyu-pi-delta} for a better agreement
with existing data. $m_0$ is the mass parameter taken as 1 GeV.

For other meson exchanges to consider, we examine the possibility
of tensor meson $a_2$ exchange with the decay of $a_2\to
\gamma\gamma$ reported in the Particle Date Group. By the vector
meson dominance the coupling $\gamma\rho a_2$ is assumed and the
coupling constant $f_{\gamma\rho a_2}=0.044$ GeV$^{-1}$ is
estimated from the constituent quark model \cite{ishida}. With the
coupling vertex $a_2N\Delta$ and the coupling constant
${f_{a_2N\Delta}\over m_{a_2}}=-3{f_{\rho N\Delta}\over m_{\rho}}$
as presented in Ref. \cite{bgyu-pi-delta} we find the contribution
of the $a_2$ exchange to be of the $10^{-2}$ order, and, hence, it
is excluded.

\begin{table}[]
\caption{\label{cc1} Physical constants for $\gamma
p\to\rho^-\Delta^{++}$. $\rho^-$ meson EM multipole moments,
$\kappa_\rho=1.01$, and $\lambda_\rho=-0.41$  are fixed in Models I
and II.  The starred multipole moments
$\lambda_{\Delta^{++}}$ and $\chi_{\Delta^{++}}$ are considered
only in Figs. \ref{fig2} (b) and \ref{fig6} (a).
EM multipoles of $\Delta^{++}$ baryon
are taken from Ref. \cite{azizi}.}
\begin{tabular}{lccccl}
                      &Model I&Model II&   \\
\hline\hline
$\kappa_{\Delta^{++}}$&$4.34$ &0      \\%
$\lambda_{\Delta^{++}}$&$6.18^*$&0    \\%
$\chi_{\Delta^{++}}$   &$12.34^*$&0    \\%
\hline
$f_{\rho^- p\Delta^{++}}$  &5.5 &8.57   \\%
$g_{\rho^- p\Delta^{++}}$  &0    &0   \\%
$h_{\rho^- p\Delta^{++}}$  &$-2$  &0     \\%
\hline
$g_{\gamma\pi\rho^\pm}$ & $0.224$&0.224   \\%
$f_{\pi^- p\Delta^{++}}$       & 1.7 &2 \\%
\hline\hline
\end{tabular}\label{tb1}
\end{table}

In the absence of other meson exchanges to contribute, the
situation is similar to the case of $\gamma N\to\rho^\pm N$
\cite{bgyu-rho}. Given the $G$-parity for photon-meson coupling
which dictates the $\rho$ exchange to change sign in accordance
with charge, we write the Reggeized production amplitude as,
\begin{eqnarray}\label{regge3}
\mathcal{M}(\rho^-)&=&-\rho\times
{1\over2}(-1+e^{-i\pi\alpha_\rho(t)})
+\pi\times 1
\end{eqnarray}
with the phases for $\rho$ and $\pi$ by
following the detailed discussion given in Ref. \cite{bgyu-rho}.

In Table \ref{tb1} we summarize the physical constants including
the complete sets for $\rho^-$ and $\Delta^{++}$ EM multipoles,
which is referred to as the Model I, hereafter. In the next
subsection, we shall discuss the minimal gauge prescription in
association with the convergence of the reaction. We refer to this
as the Model II, and list the relevant coupling constants in
advance for comparison.
\\

\begin{figure}[]
\vspace{.5cm}
\includegraphics[width=8cm]{fig2.eps}
\caption{Total cross section for $\gamma p\to \rho^-\Delta^{++}$
in the Model I. (a): The solid curve shows the cross section from
the full calculation, but without $\lambda_{\Delta^{++}}$ and
$\chi_{\Delta^{++}}$ in Table \ref{tb1} for the convergence of
solid curve up to $E_\gamma\approx$ 8 GeV.
The blue dotted curve results from the case of the approximation
$\Pi^\Delta_{\mu\nu}\approx -g_{\mu\nu}$. The
red dashed curve shows the divergence of the cross section
when the coupling $h_{\rho^- p\Delta^{++}}=0$. The dash-dotted curve
represents the contribution of gauge invariant ${\cal M}_\rho$ exchange in
Eq. (\ref{rho-regge}). (b): Roles of the $\Delta^{++}$ baryon EM
multipole moments are exhibited in the total cross section. The
blue dash-dot-dotted curve shows the case with the $\Delta^{++}$
EM multipole moments fully considered. The red dash-dash-dotted curve
results from the reaction with only the
charge coupling in the $\gamma\Delta\Delta$ vertex. Data are taken from Refs.
\cite{barber,struczinski,eisenberg,nelson}.} \label{fig2}
\end{figure}

In Fig. \ref{fig2} we present the predictions of the Model I for
the total cross section with the trajectories
\begin{eqnarray}
&&\alpha_\rho(t)=0.9(t-m_\rho^2)+1\,,\\
&&\alpha_\pi(t)=0.7(t-m_\pi^2)\,,
\end{eqnarray}
chosen consistent with $\gamma N\to\rho^\pm N$. In both panels (a)
and (b) the solid curve represents the total cross section
resulting from the physical constants fully considered in Table
\ref{tb1}, but, $\lambda_{\Delta^{++}}$ and $\chi_{\Delta^{++}}$
are turned off for a convergence of the reaction up to
$E_\gamma\approx$ 8 GeV. Though $\kappa_p$, and $\kappa_\rho$ and
$\lambda_\rho$ are turned on in the calculation, they do not alter
the results illustrated in Fig. \ref{fig2} significantly, even if
they are off. Rather, the role of $\kappa_{\Delta^{++}}$ in (b) is
of more significance within the region convergent below
$E_\gamma\approx4$ GeV, as one can expect from the magnetic nature
of the $\Delta$. Furthermore, the cross section is sensitive to
higher multipoles $\lambda_{\Delta^{++}}$ and $\chi_{\Delta^{++}}$
as well as to the strong coupling $h_{\rho^- p\Delta^{++}}$, as shown by
the divergence above $E_\gamma\approx4$ GeV. Thus, the interaction
of $\Delta^{++}$ via electromagnetic and strong couplings plays a
leading role in the lower region of the reaction. Based on this,
the case of the approximation $\Pi^\Delta_{\mu\nu}\approx
-g_{\mu\nu}$ for the $\Delta$-propagation is tested in (a)
and found to be invalid, as shown by the dotted curve.

Together with the previous result in the $\gamma
p\to\pi^-\Delta^{++}$ below $E_\gamma\approx2$ GeV
\cite{bgyu-pi-delta2}, it is interesting to observe that the
present reaction within the Regge framework for the $t$-channel
exchanges yields a positive aspect as to the study of EM
properties of $\Delta^{++}$ in the region below $E_\gamma\approx4$
GeV. In comparison to the  $\gamma p\to \pi^-\Delta^{++}$ process
where the cross section diverges $E_\gamma\simeq 1.7$ GeV above
\cite{bgyu-pi-delta2}, such a suppression of the divergence up to
8 GeV in the present process might be due to the coupling constant
$f_{\rho^- p\Delta^{++}}=5.5/m_\rho$ weaker than the $f_{\pi^-
p\Delta^{++}}=1.7/m_\pi$ in the $u$-channel $\Delta^{++}$ in the
present reaction process.
In the scheme where the spin-3/2 polarization tensor
$\Pi^\Delta_{\mu\nu}$ is fully considered for the $\Delta^{++}$
propagation in the $u$-channel, we  note that the cross section of
the Model I as shown by the solid line is valid up to
$E_\gamma\simeq8$ GeV with the desired convergence. However, such
a convergence is not persistent due to the $u$-channel divergence
of the $\Delta^{++}$ involved in Eq. (\ref{rho-regge}), though
gauge invariance is preserved.

In the next, let us consider the minimal gauge prescription for
the reaction amplitude in order for the convergence
of the reaction at high energies.

\subsection{The minimal gauge prescription: Model II}

At high energies far from the $\Delta$ resonance region, only the
single pion exchange in the $t$-channel would dominate over the
reaction process via the peripheral scattering. Therefore, in this
region, the magnetic interactions of baryons coupling to the
transverse photon field are expected to be suppressed by a factor
of $1/c$ in comparison to their charge couplings, and, hence,
negligible in the gauge invariance for the $\pi$ exchange in the
sense of minimal prescription.

More specifically, we apply a gauge prescription which introduces nucleon and $\Delta$
poles in Eq. (\ref{born}) in the minimal way, i.e., in the sense
that only the charge coupling electric Born terms in $s$- and
$u$-channels are indispensable to preserve gauge invariance of the
$t$-channel $\rho$ exchange \cite{bgyu-pi-delta}. In actual, this
is to simply remove all the terms in Eq. (\ref{born})
gauge-invariant by themselves by redundancy, which is easily
checked by replacing $\epsilon$ with $k$ and using the on-shell
condition $\overline{u}_\lambda(p')\gamma^\lambda=0$ in the EM
vertices Eqs. (\ref{gammann}) and (\ref{gamma4}).
Then, similar to the $\pi\Delta$ case as discussed in Refs.
\cite{bgyu-pi-delta,bgyu-pi-delta2}, the $u$-channel $\Delta$-pole
as given in Eq. (\ref{delta-min}) below is obtained from the
advantage of an antisymmetric form of the $u$-channel amplitude in
Eq. (12) of Ref. \cite{bgyu-pi-delta}, for instance, with help of
the term $-\epsilon^\lambda\gamma^\sigma$ in the antisymmetric
charge coupling $\gamma\Delta\Delta$ vertex in Eq. (\ref{gamma4}).

The minimal gauge-invariant $\rho$ exchange in Eq. (\ref{born}) is
now expressed as,
\begin{eqnarray}
&&{M}^{\lambda\nu\mu}_{s(N)}= {f_{\rho N\Delta}\over m_\rho}\left(
q^\lambda\gamma^\nu - \rlap{/}q g^{\lambda\nu}
\right)\gamma_5\frac{2p^\mu}{s-M^{2}_{N}}e_N,\label{nucleon-min}\\
&& {M}^{\lambda\nu\mu}_{t(\rho)}
=\Gamma^{\nu\mu\alpha}_{\gamma\rho\rho}(q,k,Q)
\frac{-g_{\alpha\beta} +Q_\alpha Q_\beta/
m^2_\rho}{t-m^{2}_{\rho}}\nonumber\\&&\hspace{1cm}\times  {f_{\rho
N\Delta}\over m_\rho}\left( Q^\beta\gamma^\lambda - \rlap{\,/}Q
g^{\beta\lambda} \right)\gamma_5,
\label{rho-min}\\
&&{M}^{\lambda\nu\mu}_{u(\Delta)}= e_\Delta\frac{2p'^\mu}
{u-M^{2}_{\Delta}} {f_{\rho N\Delta}\over m_\rho}\left(
q^\lambda\gamma^\nu - \rlap{/}q g^{\lambda\nu}
\right)\gamma_5\label{delta-min}
\end{eqnarray}
with the contact interaction term unaltered in Eq.
(\ref{contact}). A few remarks are in order: In the minimal gauge
we neglect the terms of $g_{\rho N\Delta}$ and $h_{\rho N\Delta}$
couplings, because there is no need to suppress the divergence of
the $\Delta$-pole. Strictly speaking, the $\gamma\rho\rho$ vertex
in Eq. (\ref{rho-min}) should also be simplified to have only the
charge coupling terms for consistency with others. Nevertheless,
we resume the original form to investigate the effects of
$\rho$ meson EM multipole moments, in particular, on the spin
polarization observables.

\begin{figure}[]
\includegraphics[width=8cm]{fig3.eps}
\caption{Total cross section for $\gamma p\to \rho^-\Delta^{++}$
from the Model I and II. (a): Contribution of the $\pi$ exchange
is depicted by the dashed curve. The contributions of the single
$\rho$ exchange in the $t$-channel and gauge invariant ${\cal
M}_\rho$ are given for comparison. (b): The cross section from the
Model II shows a good convergence at high energy. The dotted curve
is the cross section from without $\rho$ meson multipole moments,
i.e., $\kappa_\rho=0$ and $\lambda_\rho=0$. The contact
interaction is shown by the dash-dash-dotted curve. Notations for
curves are the same as in the panel (a). Data are taken from Refs.
\cite{barber,struczinski,eisenberg,nelson}.} \label{fig3}
\end{figure}

\begin{figure}[]
\vspace{.5cm}
\includegraphics[width=8cm]{fig4.eps}
\caption{Differential cross sections versus $-t'$ for $\gamma p\to
\rho^-\Delta^{++}$ in the Model II.
Contributions of $\pi$ and gauge invariant ${\cal M}_\rho$ exchanges
are shown in the dashed and dash-dotted curves at $E_\gamma=3.15$
GeV. The cross section from the Model I (dotted) is
presented for comparison. Data are taken from Refs.
\cite{barber,abramson}.} \label{fig4}
\end{figure}

Figure \ref{fig3} shows a comparison of the cross sections between
Model I and II with the contributions of meson exchanges therein.
Given the Model I for comparison we refer to it without
$\lambda_\Delta$ and $\chi_\Delta$ in what follows. In Fig.
\ref{fig3} (b) we demonstrate a good convergence of the cross
section at high energy by the dominant role of the $\pi$ exchange
over the $\rho$ in the minimal gauge. In order to compensate for
the absence of $\Delta^{++}$ EM multipole moments from the Model
II we choose the coupling constants $f_{\pi^- p\Delta^{++}}=2$ and
$f_{\rho^- p\Delta^{++}}=8.57$ which were adopted in the $\gamma
p\to\pi^-\Delta^{++}$ process \cite{bgyu-pi-delta}. The effect of
$\rho$ meson EM multipole moments $\kappa_\rho$ and $\lambda_\rho$
is illustrated in the total cross section by the dotted curve in
(b). The contribution of contact interaction term is presented to
compare with the case of $\gamma p\to\pi^-\Delta^{++}$, in which
case it plays the leading role in the minimal gauge
\cite{bgyu-pi-delta2}.

Differential cross sections and density matrix elements for the
unpolarized process are given in Figs. \ref{fig4} and \ref{fig5},
respectively, with respect to the modified
$t$-channel momentum squared $-t'(=t_0-t)$. $t_0$ is the value at
$\theta=0$.
The contributions of $\pi$ and $\rho$ exchanges are
shown at $E_\gamma=3.15$ GeV. The difference of the model
predictions between I and II is also presented for comparison. Both the
models predict the angular distribution of cross sections in fair
agreement with data and the observed dip at $-t\approx 0.4$
GeV$^2$ is reproduced by the nonsense-wrong-signature-zero at
$\alpha_\rho(t)=0$ from the exchange-nondegenerate phase taken for
the $\rho$ exchange.

Figure \ref{fig5} presents density matrix elements calculated
in the Gottfried-Jackson (G.-J.) frame, where the formulation of
density matrix elements for $\rho\to\pi\pi$ decay with the
$\Delta$ baryon in the final state is developed by following the
conventions and definitions of Ref. \cite{schilling}. Due to the
exchange-nondegenerate phase of the $\rho$ exchange such an
oscillatory behavior as shown in $t$-dependence of the density
matrix elements is described to a degree.

\begin{figure}[]
\includegraphics[width=8cm]{fig5.eps}
\caption{Density matrix elements $\rho^0_{\lambda\lambda'}$ versus $-t'$
at the G.-J. frame for the $\rho\to\pi\pi$ decay in the unpolarized
process $\gamma p\to \rho^-\Delta^{++}$. Notations for the curves
are the same as in Fig. \ref{fig4}. Data are taken from Refs.
\cite{barber,abramson}.} \label{fig5}
\end{figure}

%

Finally, from the magnetic nature of both the $\Delta$ and $\rho$,
when interacting with photon we expect that
the angle dependence of differential cross sections and photon polarization
asymmetry $\Sigma$ are good to observe the roles
of the magnetic dipole moments of $\Delta$ and $\rho$ in the
reaction.

\begin{figure}[]
\vspace{.5cm}
\includegraphics[width=8cm]{fig6.eps}
\caption{Differential cross sections versus angle $\theta$ for $\gamma p\to
\rho^-\Delta^{++}$.
Roles of $\Delta^{++}$ electromagnetic multipole moments are shown in (a)
and those of $\rho^-$ in (b). The sensitivity of cross section to
$\Delta^{++}$ multipole moments at backward angles shows the
characteristic $u$-channel propagation of $\Delta^{++}$.
The $t$-channel $\rho^-$ exchange exhibits a strong forward peak in (b)
with the contribution of electromagnetic multipoles
denoted by solid and dotted curves.  } \label{fig6}
\end{figure}

\begin{figure}[]
\vspace{1cm}
\includegraphics[width=8cm]{fig7.eps}
\caption{Sensitivity of photon polarization asymmetry $\Sigma$ to
magnetic dipole moments of $\Delta$ and $\rho$ with respect to
angle and to energy.
The cases of $\Sigma$ with and without $\kappa_{\Delta}$ are presented
in the upper row (a) and (b) with
$\lambda_\Delta$ and $\chi_\Delta$ set to be zero in the Model I.
The lower row (c) and (d) are the cases of the $\rho$ with and without
$(\kappa_\rho,\,\lambda_\rho)$ in the Model II.  The dashed curves
represent the dependence of $\Sigma$ on different model predictions
for $\kappa_\Delta$ and $\kappa_\rho$.} \label{fig7}
\end{figure}

Figure \ref{fig6} shows the sensitivity of the differential cross
section to the electromagnetic multipole moments of $\Delta^{++}$
and $\rho^-$. It is worth noting that the role of $\Delta^{++}$
electromagnetic multipoles is apparent at backward angles, because
of the $u$-channel $\Delta^{++}$ pole. In contrast, the
contribution of electromagnetic multipole moments of $\rho^-$ in
the $t$-channel exchange can be observed at forward angles with a
sharp enhancement of cross section reaching out to about 30
$\mu$b/str.

Figure \ref{fig7} shows the prediction for the dependence of
$\Sigma$ on angle $-t'$ in (a) and energy $E_\gamma$ in (b) in
cases of $\Delta$ baryon with and without $\kappa_\Delta$,
respectively. Likewise, the cases of $\rho$ meson with and without
EM multipole moments $(\kappa_\rho,\,\lambda_\rho)$ are given in
the lower row panels (c) and (d), respectively. Although given at
a specific energy and angle fixed around $E_\gamma=4$ GeV and
$\theta=60^\circ$ for illustration purpose, it is worth finding
that both the magnetic dipole moments of $\Delta$ and $\rho$ show
a tendency of the maximal role in the region $E_\gamma\approx4$
and 5 GeV in common. Such a similarity is found  in (a) and (c)
where the contributions of their roles become large in the  large
$-t$ equally. Furthermore, the role of $\kappa_\rho$ is more
conspicuous than that of $\kappa_\Delta$ in the spin polarization
observables just as in $\Sigma$, even though the former is less
significant in the energy  and angle dependences of the cross
sections as presented in Figs. \ref{fig3} (b) and \ref{fig6} (b).
The uncertainties in electromagnetic multipole moments of $\rho^-$
and $\Delta^{++}$ may be, at the present stage, due to the
difference of $\kappa_\Delta$ and $\kappa_\rho$ between various
model predictions. The dashed curves correspond to the cases of
different values, i.e., for $\kappa_{\Delta^{++}}=5.31$ and
$(\kappa_\rho=0.92,\,\lambda_\rho=-0.043)$ which are taken from
Refs. \cite{chiang} and \cite{crji}, respectively. The results do
not significantly alter the $\Sigma$, as presented in Fig.
\ref{fig7}.

\section{Summary and conclusions}

To summarize, we have investigated photoproduction $\gamma
p\to\rho^-\Delta^{++}$ process with particular interest in the
role of the $\rho^-$ and $\Delta^{++}$ baryon EM multipole moments as
well as the convergence of the cross section at high energies. In
order to deal with the divergence of the $u$-channel
$\Delta$-pole, we considered two sort of gauge prescriptions; one
is based on the conventional approach to gauge invariance usually
adopted in most hadron models, and the other is constrained by the
minimal scheme of gauge invariance which requires only the
longitudinal components of proton, $\rho^-$ and $\Delta^{++}$ coupling to
photon. We examined the energy dependence of the cross section up
to $E_\gamma=10$ GeV in the former, and 16 GeV in the latter
approach, respectively. The important findings in the present work
are as follows: The strong peak observed in the total cross
section near threshold is due to the predominance of the $\pi$
exchange over the $\rho$ as well as the contact interaction term.
This is quite contrasting to the case of $\gamma p\to
\pi^-\Delta^{++}$ in the minimal gauge where the contact
interaction term plays the leading role over the $\pi$. The steep
decrease of the total cross section toward high energy is well
reproduced by the longitudinal component of $\pi+\rho$ exchanges solely.
The exchange-nondegenerate phase assigned to the $\rho$ exchange
is consistent with the dip structure observed in the differential
cross section as well as the oscillatory behavior of the density
matrix elements.

Therefore, both the former and the latter approaches we referred
to as the Model I and II, respectively, can serve to our
understanding of the reaction mechanism over the baryon resonance
region without any counter terms, or cutoff functions to control
the divergence. In particular, the Model I has the advantage of
investigating the EM multipole moments of $\Delta$ baryon in the
region below $E_\gamma\simeq4$ GeV, while the Model II provides a
reliable base to explore the features of $\rho\Delta$
photoproduction at high energies without divergence.

As aforementioned, we wish to remark that the present framework
could provide a basic tool to access $\gamma p\to \pi_1\Delta$ in
a straightforward manner with the unknown physical constants
explored by the forthcoming data from high energy photon-beam
experiments such as the GlueX at the Jefferson Lab.

\section*{Acknowledgment}
This work was supported by the National Research Foundation of
Korea Grant No. NRF-2017R1A2B4010117.

\appendix
\section{ Divergence of $t$-channel $\rho$ exchange}

For illustration purpose, we present a short discussion on
the divergence of the $\rho$ exchange in the $\gamma p\to \rho^+p$
and $\gamma p\to\rho^-\Delta^{++}$
photoproductions with the term proportional to
$Q^2/m_\rho^2$ in the $t$-channel propagation.

In the former reaction
the $t$-channel $\rho$ exchange in the Born term \cite{bgyu-rho}
is written as
\begin{eqnarray}
M_t(\rho)=\Gamma^{\nu\mu\alpha}_{\gamma\rho\rho}{1\over t-m^2_\rho}
\left(-g_\rho^v\gamma_\alpha-{g_\rho^t\over4M}[\gamma_\alpha,\,{\rlap{\,/}Q}]
+g_\rho^v{Q_\alpha\over m_\rho^2}{\rlap{\,/}Q} \right).\nonumber \\
\end{eqnarray}
Thus, the term $g_\rho^v{Q_\alpha\over m_\rho^2}{\rlap{\,/}Q}$ is, in actual,
vanishing for the equal mass process, i.e., $\rlap{\,/}Q=M_p-M_n$ for on-mass shell,
which does not give rise to a divergence.

Likewise, the $\rho^-$ exchange in the Born term in Eq. (\ref{rho}) with
the leading term in the $\rho N\Delta$ coupling in Eq. (\ref{rhondelta}),
for instance, can be reduced to
\begin{eqnarray}\label{div2}
M_t(\rho)=\Gamma^{\nu\mu\alpha}_{\gamma\rho\rho}{1\over t-m^2_\rho}{f_{\rho N\Delta}
\over m_\rho}\left(g_\alpha^\lambda-{Q_\alpha Q^\lambda\over m_\rho^2}\right)
{\rlap{\,/}Q},
\end{eqnarray}
if $Q^2=m_\rho^2$ is assumed for on-shell. The contribution
proportional to the term  $Q^2/m_\rho^2$ in this case is not vanishing,
but constrained by the mass difference, $M_p-M_\Delta$.
The result in Eq. (\ref{div2}) with the overall factor $\rlap{\,/}Q$
also shows a reason for the contribution of the single
$\rho$ exchange much smaller than that of ${\cal M}_\rho$ in
Fig. \ref{fig3} (a).

In the similar fashion,
the divergence of the $\Delta$ propagation due to the term $p_\Delta^2/M_\Delta^2$
in the $u$-channel can be traced out.


\end{document}